\documentclass[aip,reprint,floatfix]{revtex4-1}
\usepackage{amsfonts,amssymb,amsmath,array}

\usepackage{dcolumn}
\usepackage{bm}

\usepackage[final]{graphicx}
\usepackage{graphicx}
\usepackage{epstopdf}
\graphicspath{{./Figures/}} 

\usepackage[usenames]{color} 

\usepackage{soul} 

\begin{document}

\title{Chiral active matter in external potentials}

\author{Lorenzo Caprini}
\email{lorenzo.caprini@gssi.it, lorenzo.caprini@hhu.de}
\affiliation{Heinrich-Heine-Universit\"at D\"usseldorf, Institut f\"ur Theoretische Physik II - Soft Matter, 
D-40225 D\"usseldorf, Germany}

\author{Hartmut L{\"o}wen}
\email{hlowen@hhu.de}
\affiliation{Heinrich-Heine-Universit\"at D\"usseldorf, Institut f\"ur Theoretische Physik II - Soft Matter, 
D-40225 D\"usseldorf, Germany}

\author{Umberto Marini Bettolo Marconi}
\affiliation{ Scuola di Scienze e Tecnologie, Universit\`a di Camerino - via Madonna delle Carceri, 62032, Camerino, Italy and
INFN Sezione di Perugia, I-06123 Perugia, Italy}

\begin{abstract}
We investigate the interplay between chirality and confinement induced by the presence of an external potential. 
For potentials having radial symmetry, the circular character of the trajectories induced by the chiral motion reduces the spatial fluctuations of the particle, thus providing an extra effective confining mechanism, that can be interpreted as a lowering of the  effective temperature.
In the case of non-radial potentials, for instance, with an elliptic shape, chirality displays a richer scenario.
Indeed, the chirality can break the parity symmetry of the potential that is always fullfilled in the non-chiral system.
The probability distribution displays a strong non-Maxwell-Boltzmann shape that emerges in cross-correlations between the two Cartesian components of the position, that vanishes in the absence of chirality or when radial symmetry of the potential is restored. These results are obtained by considering two popular models in active matter, i.e. chiral Active Brownian particles and chiral active Ornstein-Uhlenbeck particles.
\end{abstract}

\maketitle

\section{Introduction}

Active matter, encompassing a wide range of self-propelled entities, has emerged as a fascinating field of study in soft matter and non-equilibrium statistical physics~\cite{marchetti2013hydrodynamics, bechinger2016active}.
Typical active systems are artificial particles, such as active colloids, active granular particles, and drones, but also living systems with biological origins, such as bacteria, sperms, and several animals.
These systems usually self-propel by virtue of internal mechanisms that convert energy to produce a net motion, through chemical reactions, cilia, flagella, and internal motors, to mention a few examples.

In several cases, the self-propelled motion is characterized by an almost straight path and a fluctuating orientation that changes stochastically without a preferential direction. This motion is induced by the breaking of the translational symmetry at the single-particle level in the body or in the swimming and running mechanism that induces a net polarity in the particle.
The physical or biological systems displaying this motion are classified as {\it linear} particles or swimmers.
This is the standard scenario for several bacteria, such as E. Coli, active colloids, such as Janus particles, or polar active granular particles.
However, in nature, several active systems show trajectories systematically rotating clockwise or counterclockwise, the so-called {\it chiral} or {\it circular} self-propelled particles~\cite{lowen2016chirality}.

The concept of chirality or handedness was introduced by Lord Kelvin more than one century
ago in reference to the circular (helical) motion produced by solid bodies with asymmetric shapes in two (three)
dimensions. Nowadays, chirality has been renewed in the field of active matter~\cite{liebchen2022chiral}, being observed for instance in proteins~\cite{loose2014bacterial}, bacteria~\cite{diluzio2005escherichia, lauga2006swimming} and sperms~\cite{riedel2005self} moving on a two-dimensional planar substrate, and L-shape artificial microswimmers~\cite{kummel2013circular}. In addition, even spherical (non-chiral) particles can show circular (chiral) trajectories due to asymmetry in their self-propulsion mechanism, as occurs in colloidal propellers in a magnetic or electrical field~\cite{zhang2020reconfigurable}, and cholesteric droplets~\cite{carenza2019rotation}. In addition, granular systems such as spinners~\cite{workamp2018symmetry, scholz2021surfactants} and Hexbug particles driven by light~\cite{siebers2023exploiting} usually display chiral motion.

Being ubiquitous in nature, the interest in chiral active matter is recently showing exponential growth in time, in different contexts ranging from the statistical properties of single-particles to collective phenomena displayed by interacting systems.
Through the introduction of simple models, the single-particle chiral active motion has been explicitly explored~\cite{wittkowski2012self, kummel2013circular} with a focus on the mean-square displacement~\cite{van2008dynamics, sevilla2016diffusion}, in a viscoelastic medium~\cite{sprenger2022active}, in the presence of pillars~\cite{van2022role} or sinusoidal channels~\cite{ao2015diffusion}.
In channel geometries, chirality is also responsible for the reduction of the accumulation near boundaries typical of active systems and for the formation of surface currents~\cite{caprini2019activechiral, fazli2021active}.
In the case of interacting systems, chirality is able to suppress the clustering typical of active particles~\cite{ma2022dynamical, semwal2022macro, sese2022microscopic, bickmann2022analytical} but induces novel phenomena, such as emergent vortices induced by the chirality~\cite{liao2018clustering, liao2021emergent} or a global traveling wave in the presence of a chemotactic alignment~\cite{liebchen2016pattern}.
Chiral active particles exhibit fascinating phenomena also in the presence of alignment interactions giving rise to pattern formation~\cite{liebchen2017collective, negi2022geometry} consisting of rotating macro-droplets~\cite{levis2018micro}, chiral self-recognition~\cite{arora2021emergent}, dynamical frustration~\cite{huang2020dynamical}, and chimera states~\cite{kruk2020traveling}.
In addition, chirality appears as a fundamental ingredient to observe the hyper-uniform phase~\cite{lei2019nonequilibrium, huang2021circular} in active matter as well as emerging odd properties~\cite{fruchart2023odd, muzzeddu2022active} for instance in the viscosity~\cite{banerjee2017odd, lou2022odd,yang2021topologically, lou2022odd}, elasticity~\cite{scheibner2020odd}, and mobility~\cite{poggioli2023odd}.
Recently, the circular motion has been also investigated in the framework of active glasses where it gives rise to a novel oscillatory caging effect entirely due to the chirality~\cite{debets2023glassy}.

Chirality could play a fundamental role in several applications due to their emerging properties, such as sorting~\cite{mijalkov2013sorting, chen2015sorting, su2019disordered, xu2022sorting} and synchronization~\cite{levis2019activity, samatas2023hydrodynamic}.
For instance, chiral microswimmers can be sorted according to their swimming properties by employing patterned microchannels with a specific chirality~\cite{mijalkov2013sorting}.
Chirality is also at the basis of the ratcheting mechanism observed in an array of obstacles~\cite{reichhardt2013dynamics} even leading to translation at fixed angles with respect to the substrate periodicity due to a periodic potential~\cite{nourhani2015guiding}.
Moreover, binary mixtures of passive and active chiral particles, as well as mixtures of chiral particles with opposite chiralities show demixing~\cite{ai2015chirality, ai2018mixing, reichhardt2019reversibility, levis2019simultaneous}.
Spontaneous demixing has been also observed experimentally in a system of active granular particles, the so-called spinners that are self-propelled because of the asymmetry of internal components of their bodies~\cite{scholz2018rotating}.

\begin{figure}[t!]
\includegraphics[width=0.8\columnwidth]{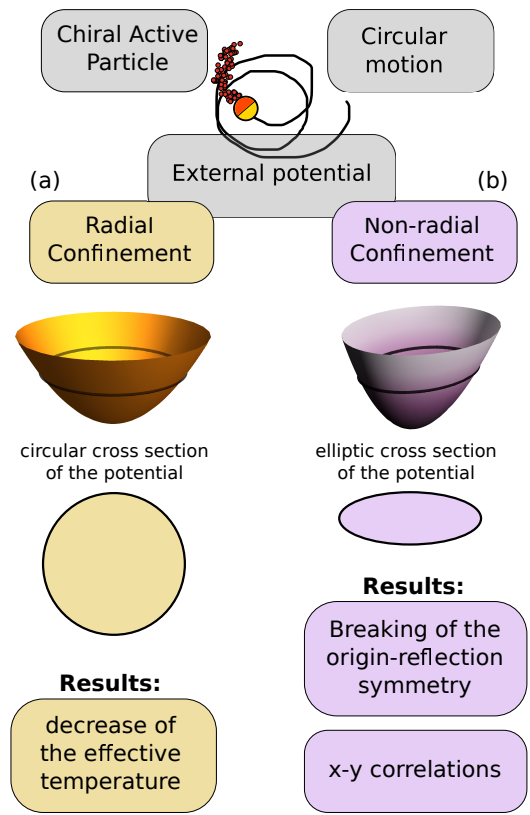}
\caption{
{\textbf{Chiral active particles in external confinement}.}
Illustration of a chiral active particle displaying circular motion.
In the presence of an external potential with radial symmetry (circular cross-section), a decrease in the effective temperature is induced by the increase of the chirality.
In contrast, for non-radial external potential with an elliptic cross-section, the chirality breaks the parity symmetry typical of the potential leading to a non-Boltzmann probability distribution with emerging correlations between different Cartesian components of the position.
}
\label{fig:fig0}
\end{figure}

Despite the recent attention on chiral active matter, the interplay between chirality and external confinement due to an external potential has been less investigated \cite{jahanshahi2017brownian} to the best of our knowledge.
Here, we focus on active chiral particles in a radial (circular) and non-radial (elliptic) potential, exploiting the influence of circular motion on the properties of the system.
In particular, we perform a numerical and analytical study based on two popular models in active matter, i.e. the chiral active Brownian particles and chiral active Ornstein-Uhlenbeck particles.
We anticipate that for a radial potential, the chirality induces only an increasing confinement in the particle's dynamics, effectively reducing the fluctuations of the systems and, thus its effective temperature (Fig.~\ref{fig:fig0}~(a)).
In contrast, in the case of non-radial potential, the chirality is able to break the parity symmetry of an elliptic potential. This is reflected, for instance, in the occurrence of strong correlations between different spatial components of the system (Fig.~\ref{fig:fig0}~(b)). This effect is uniquely based on the interplay between chirality and spatial asymmetry of the potential.

The paper is structured as follows:
in Sec.~\ref{sec:Model}, we introduce and discuss the models, i.e. chiral active Brownian particles and chiral active Ornstein-Uhlenbeck particles, employed to perform the numerical and analytical study.
The dynamics in the radial and non-radial potentials are analyzed in Sec.~\ref{sec:radial} and Sec.~\ref{sec:nonradial}, respectively.
 We summarize the results and report a conclusive discussion in the final section~\ref{sec:conclusions}.
Finally, for the sake of completeness but also to render the presentation lighter, we reported
in an appendix the derivation of the Fokker-Planck equation governing the evolution of the probability distribution function
of the chiral active model together with a pair of simple illustrative cases.

\section{Model} \label{sec:Model}

Active particles in the overdamped regime are described by the following dynamics for the particle position $\mathbf{x}$:
\begin{equation}
\gamma\dot{\mathbf{x}}= \mathbf{F}(\mathbf{x})+ {\gamma}\sqrt{2 D_t} \mathbf{w} + \gamma v_0\mathbf{n} \,,
\label{eq_t1}
\end{equation}
where $\mathbf{w}$ is a Brownian white noise with unit variance and zero average accounting for the random collisions with the particle of the solvent.
The coefficient $\gamma$ is the friction coefficient due to the solvent, while $D_t$ is the translational diffusion coefficient of the system. 
The term $\mathbf{F}(\mathbf{x})$ is the external force due to a potential $U(\mathbf{x})$, such that $\mathbf{F}=-\nabla U$.
The last force term in Eq. \eqref{eq_t1}, namely $v_0 \gamma \mathbf{n}$, known as active force, describes at a coarse-grained level the chemical, biological or physical mechanism responsible for the self-propulsion.
The constant $v_0$ provides a velocity scale to the dynamics and it is often referred to in the literature as swim velocity, while the vector $\mathbf{n}$ is a stochastic process with unit variance whose properties and dynamics determine the active model considered.
$\mathbf{n}$ is an additional degree of freedom that is absent for equilibrium systems where $v_0=0$.
Despite the generality of Eq.~\eqref{eq_t1}, for simplicity, we restrict ourselves to two spatial dimensions.

\subsection{Chiral active Brownian particles (ABPs).}
In the ABP dynamics~\cite{buttinoni2013dynamical, solon2015pressure, shaebani2020computational, caporusso2020motility, caprini2023flocking} independently of the chirality, the term $\mathbf{n}$ is a unit vector, such that $|\mathbf{n}|=1$, usually associated with the orientation of the active particle.
Since the modulus of $\mathbf{n}$ is unitary, the dynamics of $\mathbf{n}$ can be conveniently expressed in polar coordinates.
In this representation, $\mathbf{n}=(\cos\theta, \sin\theta)$, where $\theta$ is the orientational angle of the active particle that evolves as a simple diffusive process:
\begin{equation}
\dot{\theta} = \sqrt{\frac{2}{\tau}}\, \xi + \omega\,,
\label{eq_thetaABP}
\end{equation}
where $\xi$ is a white noise with unit variance and zero average and the typical time $\tau$ can be identified with the persistence time
induced by the rotational diffusion coefficient $D_\text{r}=1/\tau$.

In the ABP dynamics, the chirality is introduced by adding an angular drift $\omega$ in Eq.~\eqref{eq_thetaABP}, which
breaks the rotational symmetry of the active force dynamics and induces a preferential rotation of the vector $\mathbf{n}$ in the clockwise or counterclockwise direction depending on the sign of $\omega$.
As a consequence, the single-particle trajectories of a chiral ABP 
tend to be circular.
The value of $|\omega|$ determines the strength of chirality: the larger $\omega$, the smaller the typical radius of the circular trajectories of a single particle, given by $v_0/\omega$.

\subsection{Chiral active Ornstein-Uhlenbeck particles (AOUPs).} 
In the AOUP dynamics~\cite{szamel2014self, martin2021statistical, maggi2015multidimensional, wittmann2018effective,  caprini2018activeescape, keta2022disordered, keta2023intermittent}, $\mathbf{n}$ is described by a two-dimensional Ornstein-Uhlenbeck process that allows both the modulus $|\mathbf{n}|$ and the orientation $\theta$ to fluctuate with related amplitudes~\cite{caprini2022parental}.
The AOUP distribution is a two-dimensional Gaussian such that each component fluctuates around a vanishing mean value with unit variance.
The resulting dynamics of the vector $\mathbf{n}$ reads:
\begin{equation}
\dot{\mathbf{n}}= - \frac{\mathbf{n}}{\tau} + \sqrt{\frac{1}{\tau}} \boldsymbol{\chi} + \omega \,\mathbf{n}\times \mathbf{z}
\end{equation}
where $\boldsymbol{\chi}$ is a two-dimensional vector of white noises with uncorrelated components having unitary variance and zero average.
Here, $\tau$ represents the persistence time of the particle trajectory, i.e. the time that the particle,
in the absence of angular drift, spends moving in the same direction before a reorientation of the active force.
In the AOUP model the diffusion coefficient due to the active force is
obtained form the relation $2D_a/\tau= v_0^2\tau$, which allows 
a simple comparison between AOUP and ABP models~\cite{caprini2022parental, caprini2019comparative}.

In the AOUP dynamics, the chirality is included by adding the force $\omega \,\mathbf{n}\times \mathbf{z}$, where $\mathbf{z}$ is the direction orthogonal to the plane of motion and the parameter $\omega$ quantifies the chirality of the particle~\cite{caprini2019activechiral}.
Such a force is always directed in the plane of motion, normal to $\mathbf{z}$,
and is orthogonal to $\mathbf{n}$, so that it rotates the self-propulsion vector in the clockwise or counterclockwise direction depending on the sign of $\omega$. 
Similarly to the chiral ABP model, the chiral AOUP dynamics displays circular trajectories. However,  in contrast with the ABP dynamics, the typical circles observed by an AOUP are characterized by a fluctuating radius, that
on average is equal to the one of the ABP and  $\approx v_0/|\omega|$.
It is worth noting that the chiral term in the AOUP dynamics is totally equivalent to the chiral term in the ABP dynamics. Indeed, the constant force $\omega \,\mathbf{n}\times \mathbf{z}$ in polar coordinate affects only the dynamics of the polar angle through a constant term equivalent to the driving angular velocity written in Eq.~\eqref{eq_thetaABP}.

\subsection{Relation between chiral AOUPs and chiral ABPs.} 
Despite the AOUP and ABP dynamics are different, both are usually employed to describe active particles and display similarities so that AOUP has been often employed to derive analytical predictions suitable to describe ABP numerical results.
The reason of this agreement lies in the fact that the two-time self-correlations of $\mathbf{n}$ of the two models are identical
with an approprate choice of parameters~\cite{farage2015effective, caprini2022parental,caprini2019activechiral}.
For both cases, we find
\begin{align}
\langle \mathbf{n}(t)\cdot\mathbf{n}(0)\rangle=e^{-\frac{t}{\tau}} \cos(\omega t) \,.
\label{eq_nncorr}
\end{align}
It is worth noting that, in Eq.~\eqref{eq_nncorr}, the chirality affects the shape of the autocorrelation by inducing oscillations.

Despite ABP and AOUP have different dynamics and are characterized by different steady-state distributions, such dynamical properties are at the basis of a plethora of similar phenomena observed for a single particle but also for interacting systems.
A comparison between the two models has been established for a single non-chiral active particle and a non-chiral active particle in a harmonic potential, while, more generally, the relation between the two models has been deepened in Ref.~\cite{caprini2022parental}.
However, the effect of chirality in the two models confined in an external potential has been poorly investigated in the literature.

\section{Chiral active particle in a radial potential} \label{sec:radial}

We start by considering chiral active particles confined by a simple harmonic potential in two dimensions, $U(\mathbf{x}) = k \mathbf{x}/2$, that exerts a linear force on the particle directed towards the origin.

Both in chiral ABP and chiral AOUP simulations, it is convenient to rescale time by the persistence time $\tau$ and the position by the persistence length $v_0 \tau$.
In this way, the chirality can be tuned by changing the dimensionless parameter $\omega \tau$, which we call reduced chirality.
The other dimensionless parameters of the simulations are the reduced stiffness of the potential $k\tau/\gamma$ and the ratio between passive and active diffusion coefficients, $D_t/(\tau v_0^2)$.
For simplicity, we set $D_t=0$ and eliminate $D_t/(\tau v_0^2)$.
Indeed, the thermal noise is orders of magnitudes smaller than the diffusion due to the active force in several experimental systems \cite{bechinger2016active}.
Finally, we set $k\tau/\gamma=1$.
The effect of this parameter has been explored in the AOUP case analytically~\cite{szamel2014self}, and in the ABP case numerically~\cite{caprini2022parental} and experimentally~\cite{buttinoni2022active} by considering an active Janus particle in an optical tweezer.
Here, we focus on the role of reduced chirality,  $\omega\tau$.

\begin{figure*}[t!]
\includegraphics[width=0.8\textwidth]{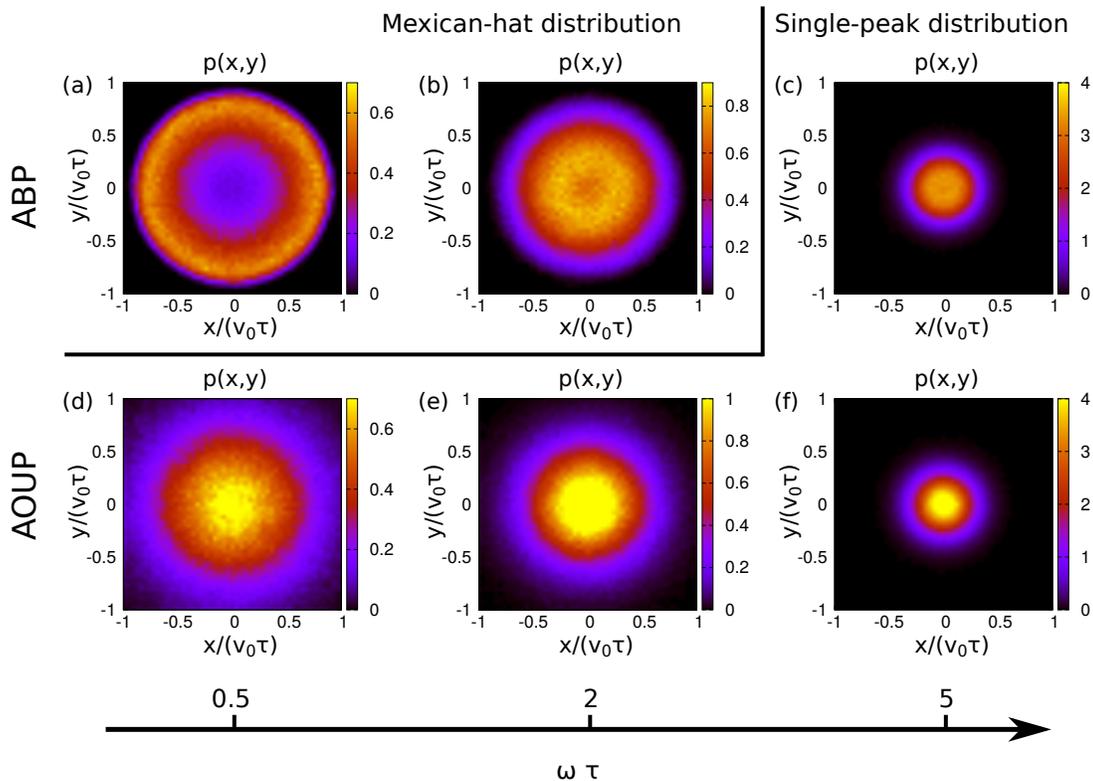}
\caption{
{\textbf{Probability distributions for an active chiral particle in a harmonic radial potential}.}
Probability distribution, $p(x,y)$, as a function of the rescaled position, $x/(v_0 \tau)$ and $y/(v_0 \tau)$, for a chiral active particle in a harmonic potential. Panels (a)-(c) are obtained by considering the ABP dynamics, while panels (d)-(f) are obtained by considering the AOUP dynamics.
The $p(x,y)$ are shown for several values of the reduced chirality, $\omega \tau$, as indicated in the figure: $\omega \tau=0.5$ (panels (a) and (d)), $\omega\tau=2$ (panels (b) and (e)), $\omega\tau=5$ (panels (c) and (f)).
The remaining parameters of the simulations are: $k \tau/\gamma=1$ and $D_t/(v_0^2 \tau)=0$.
}
\label{fig:fig1}
\end{figure*}

Active particles in radial potentials~\cite{takatori2016acoustic, dauchot2019dynamics, caprini2019activity, hennes2014self, rana2019tuning, marini2017pressure, baldovin2022control} have been widely investigated in the absence of chirality for which we summarize the results:
the AOUP dynamics in a harmonic potential can be solved exactly~\cite{szamel2014self, das2018confined, woillez2020active, caprini2020inertial, nguyen2021active}, being fully linear, and is described by a multivariate Gaussian distribution in $\mathbf{x}$ and $\mathbf{n}$. As a consequence, the density $p(\mathbf{x})$ of the system is still Gaussian and the active force affects the distribution by changing its effective temperature only~\cite{szamel2014self, fodor2016far, maggi2017memory, caprini2022active}.
The ABP dynamics in harmonic potential has been exactly solved only recently~\cite{malakar2020steady, caraglio2022analytic} and leads to a more intriguing scenario~\cite{pototsky2012active, basu2019long}. While in the small persistence regime, (small $\tau$ or large $D_r$), the density is Gaussian~\cite{caprini2022parental} and similar to the one of the AOUP, in the large persistence regime, ABPs accumulate far from the potential minimum, as confirmed experimentally by active colloids~\cite{takatori2016acoustic, buttinoni2022active}, roughly at the distance where the active force balances the potential force, i.e. at $|\mathbf{x}| \approx v_0\gamma/k$ . As a result, the two-dimensional density in the plane of motion is characterized by a Mexican-hat shape while the density, projected onto a single coordinate, displays bimodality. 
The results observed in the ABP are reminiscent to those originally obtained of considering Run\&Tumble particles~\cite{tailleur2009sedimentation, solon2015active, smith2022exact}.

\subsection{Spatial distribution}
To investigate the role of chirality, we plot the probability distribution $p(x,y)$ in the plane of motion for three representative values of the reduced chirality, $\omega \tau$.
This analysis is performed both for the ABP (Fig.~\ref{fig:fig1}~(a)-(c) ) and AOUP (Fig.~\ref{fig:fig1}~(d)-(f)) models.

In the chiral AOUP case, the system is linear and, as a consequence, $p(x,y)$ is a Gaussian centered at the origin in both spatial directions, independently of the value of $\omega \tau$.
The increase of the chirality induces a stronger confinement of the particle as if the potential was stiffer or
the dynamics governed by a lower effective temperature.
Indeed, the system is described by the following $p(x,y)$
\begin{equation}
\label{eq:p(x,y)theory}
p(x,y)=\mathcal{N} \exp\Bigl( - \frac{k (x^2+y^2}{2 T_{eff}} \Bigr)
\end{equation}
with effective temperature (in units of Boltzmann constant, $k_B=1$)
\begin{equation}
\label{eq:theory_harmonic_kinetictemperature}
T_{eff}=\frac{\langle x^2\rangle}{k}=\frac{1+\frac{\tau}{\gamma}k}{ (1+\frac{\tau}{\gamma} k)^2+\omega^2\tau^2 } \,\tau \gamma v_0^2 \,.
\end{equation}
The theoretical results~\eqref{eq:p(x,y)theory} and~\eqref{eq:theory_harmonic_kinetictemperature} are derived in Appendix \ref{sec:appendixB}, while the general method is described in Appendix~\ref{sec:appendixB}.
The effective temperature $T_{eff}$ is consistent with the expression for $\omega \tau \ll 1$, which a decrease as $\tau \to 0$ and an increase proportional to $v_0^2$.
The effect of chirality $\omega$ manifests itself as a decrease of the effective temperature, consistently with Figs.~\ref{fig:fig1}~(a),~(b), and~(c).

As  expected, the ABP case is richer:
for small values of $\omega\tau \lesssim 1$, chiral ABPs accumulate at a finite distance from the minimum of the potential (Fig.~\ref{fig:fig1}~(a)) as already observed in the absence of chirality.
The distribution displays the typical Mexican-hat shape, i.e. the particles accumulate on a ring roughly at distance $\approx v_0 / k$ from the origin. In this regime, the increase of the chirality broadens the width of the ring.
The tendency of particles to rotate (on average) in a clockwise (counterclockwise) direction hinders the ability of the particles to accumulate out of the minimum: a particle accumulated at a radial distance $\approx v_0 \gamma/k$ could change the direction of the active due to the rotation induced by the chirality.
For larger values of $\omega \tau \sim 1$, the rotations of the particles are stronger and characterized by a smaller radius of the circle. Thus, the accumulation is observed at a position much closer to the minimum of the potential  with respect to the previous case (Fig.~\ref{fig:fig1}~(b)): particles cannot reach the position $v_0\gamma /k$ before the chirality turns the direction of the active force before the particles arrive at this position.
Finally, the accumulation is completely suppressed for $\omega\tau \gtrsim 1$, when the particle simply performs small circular trajectories around the minimum of the potential.
In the latter regime (Fig.~\ref{fig:fig1}~(c)), $p(x,y)$ is again peaked at the origin and the effect of chirality can be mapped again onto an effective temperature.
This occurs because the radius of the circular trajectory, namely $v_0/\omega$ is smaller than the typical distance at which particles accumulate $v_0/k$. As a consequence, particles' ability to climb on the potential is contrasted by their tendency to spin and perform circular trajectories around the potential minimum.

\begin{figure}[t!]
\includegraphics[width=\columnwidth]{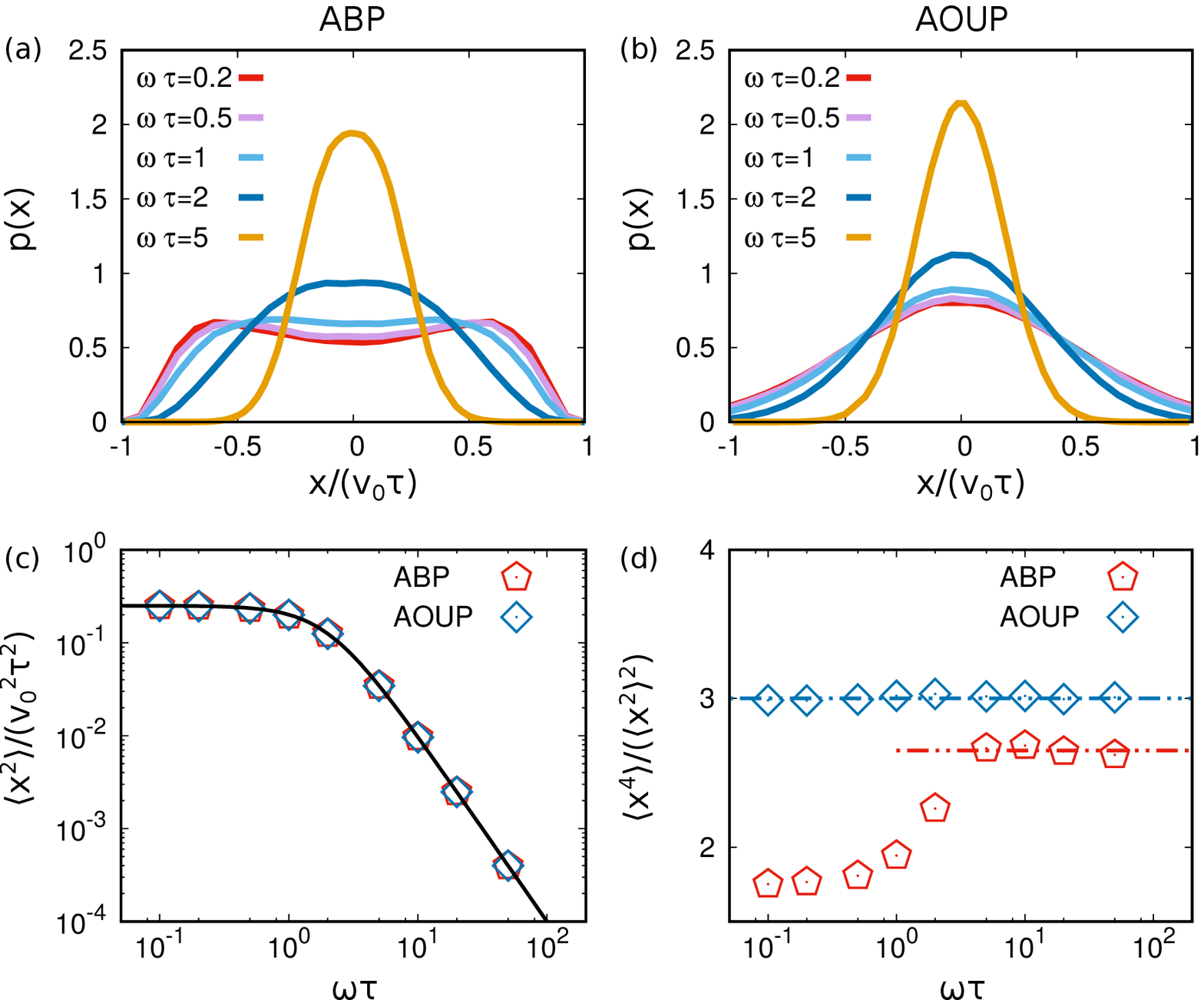}
\caption{
{\textbf{Longitudinal density and its moment for an active chiral particle in a harmonic radial potential}.}
Panels (a) and (b): density distribution $p(x)$ projected onto the $x$ axis as a function of the rescaled position $x/(v_0\tau)$ for several values of the reduced chirality. Panels (a) and (b) are obtained by considering the ABP and the AOUP dynamics, respectively.
Panel (c): variance of the distribution $\langle x^2 \rangle$ as a function of $\omega\tau$.
Panel (d): Kurtosis of the distribution $\langle x^4 \rangle / \langle x^2 \rangle^2$ as a function of $\omega \tau$.
Both in panels (c) and (d), ABP and AOUP are represented by red and blue symbols.
The black solid line in panel (c) represents the theoretical prediction, Eq.\eqref{eq:theory_harmonic_kinetictemperature}, the dashed blue line in panel (d) marks the value corresponding to the Gaussian prediction and finally, the red dashed line is an eye-guide marking the asymptotic value obtained by the Kurtosis of the ABP.
The remaining parameters of the simulations are: $k \tau/\gamma=1$ and $D_t/(v_0^2 \tau)=0$.
}
\label{fig:fig2}
\end{figure}

\subsection{Projected density and moments of the distribution}

In Fig. \ref{fig:fig2} the spatial density, $p_1(x)$, projected onto a single spatial component are plotted for several values of reduced chirality $\omega \tau$.
As expected, the ABP case (Fig.~\ref{fig:fig2}~(a)) is richer than the AOUP case (Fig.~\ref{fig:fig2}~(b)).
The latter is characterized by a Gaussian $p_1(x)$, whose variance varies with $\omega \tau$, while the former shows a transition from a bimodal distribution (characterized by two lateral peaks) to a unimodal distribution, when $\omega \tau \gtrsim 1$.
We consider the moment of this distribution both for ABP and AOUP cases.
By symmetry, the first moment is zero, while
in both models, the variance $\langle x^2 \rangle$ of $p(x)$ displays a monotonic decrease with $\omega \tau$ starting at $\omega \tau \sim 1$.
For the variance of the distribution, both AOUP and ABP dynamics show consistent results.
Finally, we study the kurtosis of the distribution $\langle x^4\rangle / \langle x^2\rangle^2$ in the AOUP and ABP to quantify the non-Gaussianity of the latter.
In the AOUP case, the kurtosis is equal to $3$ being the model Gaussian, whereas
in the ABP, the kurtosis is always smaller than $3$ as a result of the non-Gaussian nature of the distribution.
As $\omega \tau$ increases the kurtosis goes from a  value  $\approx 2$ (when $p_1(x)$ is bimodal) to a large asymptotic value sightly  smaller than $3$ (where $p_1(x)$ is unimodal).
This implies that the chirality reduces the non-Gaussianity of the distribution but that the unimodal $p_1(x)$ observed for larger $\omega \tau$ is still non-Gaussian.

\section{Chiral active particle in a non-radial potential}\label{sec:nonradial}

In this section, we investigate the dynamics of an active chiral particle in a potential that breaks the rotational symmetry of the system. We consider a harmonic potential with an elliptic shape: $U(x,y)=\frac{1}{2}(k_x x^2 + k_y y^2)$.
Such a potential introduces an additional dimensionless parameter, $k_y/k_x $, which quantifies the asymmetry of the potential and chose $k_y/k_x =3$.
The remaining dimensionless parameters are $k_y \tau/\gamma =1$ and $D_t/(\tau v_0^2)=0$.
Here, again we vary the reduced chirality $\omega \tau$ to study the interplay between chirality and asymmetry of the potential.

The asymmetry between the two orthogonal directions in the corresponding equilibrium system 
would be fully described by the Maxwell-Boltzmann distribution: particles fluctuate around the origin and explore larger regions of space along the direction where the potential gradient is weaker.
The generalization to non-chiral active particles is rather straightforward both for AOUP and ABP and does not present significant changes with respect to the symmetric case.
Indeed, the non-chiral AOUP in the potential $U(x,y)$ is characterized by a Gaussian distribution similar to the equilibrium case, while the non-chiral ABP, displays accumulation away from the minimum on an ellipsoidal domain rather than a circular one.
Intuitively, the accumulation along the more confined direction will be stronger.

\subsection{Spatial distribution and cross-correlations}

\begin{figure*}[t!]
\includegraphics[width=0.95\textwidth]{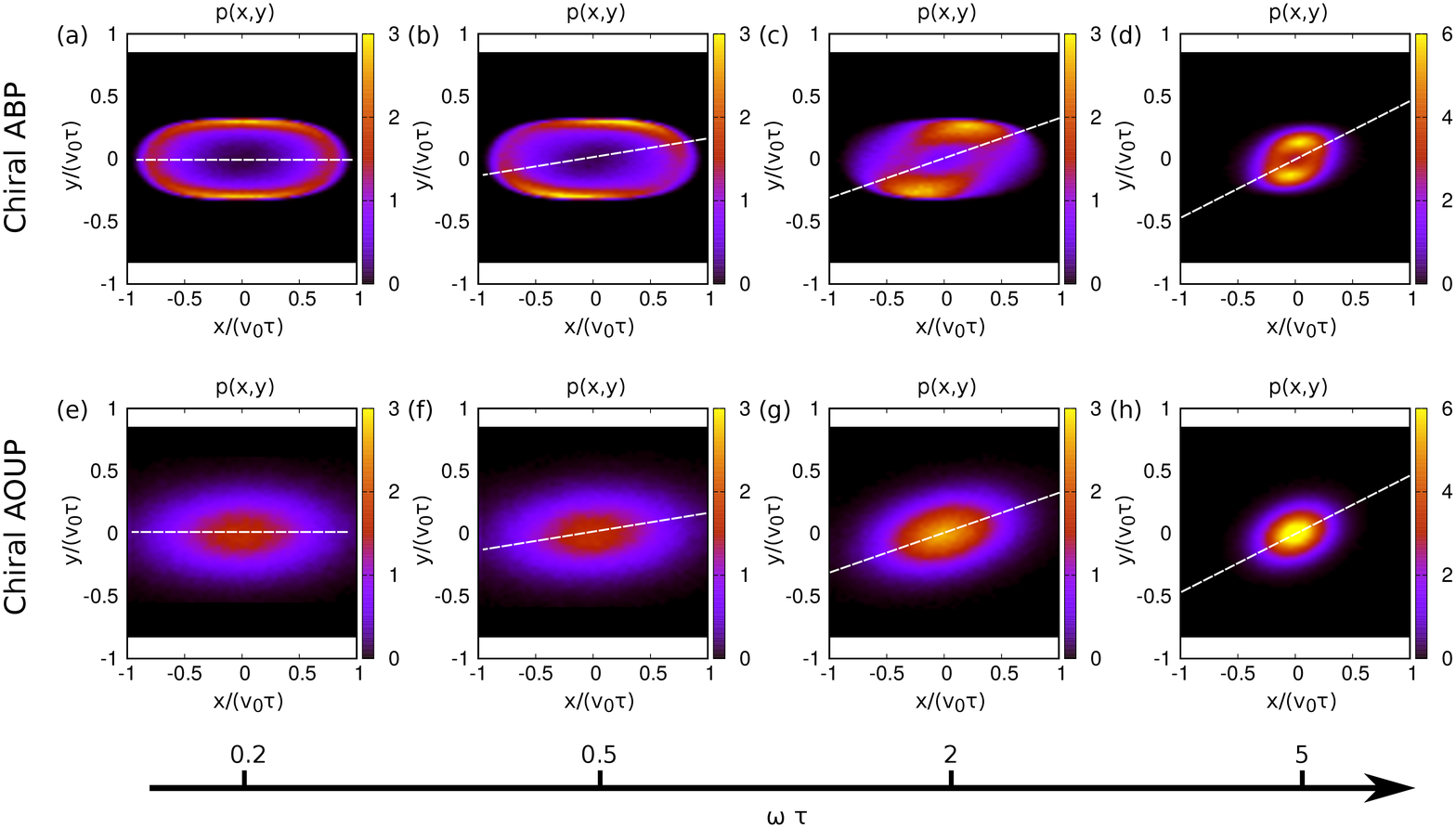}
\caption{
{\textbf{Probability distributions for an active chiral particle in a harmonic elliptic potential}.}
Probability distribution, $p(x,y)$, as a function of the rescaled position, $x/(v_0 \tau)$ and $y/(v_0 \tau)$, for a chiral active particle in a harmonic potential. Panels (a)-(d) are obtained by considering the ABP dynamics, while panels (e)-(h) are obtained by considering the AOUP dynamics.
The $p(x,y)$ are shown for several values of the reduced chirality, $\omega \tau$, as indicated in the figure: $\omega \tau=0.2$ (panels (a) and (e)), $\omega\tau=0.5$ (panels (b) and (f)), $\omega\tau=2$ (panels (c) and (g)), $\omega\tau=5$ (panels (d) and (h)).
The remaining parameters of the simulations are: $k_y/k_x=3$, $k \tau/\gamma=1$, and $D_t/(v_0^2 \tau)=0$.
}
\label{fig:fig3}
\end{figure*}

The role of chirality in a harmonic elliptic potential is analyzed by studying the two-dimensional density distribution $p(x,y)$. The analysis is performed both for ABP and AOUP dynamics and for several values of the reduced chirality $\omega \tau$ (Fig.~\ref{fig:fig3}).

In the AOUP case (Fig.~\ref{fig:fig3}~(e)-(h)), $p(x,y)$ displays a Gaussian shape, i.e. particles preferentially explore the spatial regions close to the origin, i.e. the minimum of the potential.
For small $\omega \tau \ll 1$ (Fig.~\ref{fig:fig3}~(e)), the findings are consistent with the non-chiral scenario: active particles explore the elliptic region around the origin and the chirality slightly decrease the spatial fluctuations as seen in the case of a radial potential.
The effect of the chirality emerges for larger values of $\omega \tau$.
As shown in Fig.~\ref{fig:fig3}~(f)-(h), the chirality  tilts the main axis of the ellipse where the particles accumulate.
As a consequence, $p(x,y)$ has a non-Maxwell-Boltzmann shape, since the distribution cannot be expressed as $p(x,y) \sim e^{-U/T_{eff}}$, with $k_B=1$.
As already remarked, this effect is absent for non-chiral AOUP, and, thus, is purely induced by the interplay between the chirality and the breaking of the radial symmetry of the confining potential.
In general, we observe that the increase of $\omega \tau$ increases the tilt angle of the ellipsoid until it reaches a saturation value that by symmetry cannot exceed $\pi/4$.
Finally, for $\omega \tau \gtrsim 1$ the chirality leads to a stronger confinement and, thus, decreases the effective temperature of the system without altering the ellipsoidal shape of the potential, as shown from Fig.~\ref{fig:fig3}~(g) to Fig.~\ref{fig:fig3}~(h).
The last observation is consistent with the finding relative to the radial potential of Sec.~\ref{sec:radial}.

The numerical results are confirmed by the expression for the probability distribution $p(x,y)$ that reads (see Appendix~\ref{sec:appendixB})
\begin{equation}
\label{eq:pxy_elliptic}
p(x,y)=C\exp \Bigl(-\frac{1}{2}\frac{\langle y^2 \rangle x^2+\langle x^2 \rangle y^2- 2\langle xy\rangle xy}{\langle x^2 \rangle\langle y^2 \rangle-\langle xy \rangle^2}\Bigr)
\end{equation}
where the variances $\langle x^2\rangle$ and $\langle y^2 \rangle$ are given by
\begin{flalign}
\label{eq:elliptic_theory_varx}
\langle x^2\rangle&=
\frac{ v_0^2\tau\gamma}{ k_x } \frac{ (1+\frac{\tau}{\gamma}k_x) }{(1+\frac{\tau}{\gamma}k_x)^2 +\Omega^2\tau^2}\\
\label{eq:elliptic_theory_vary}
\langle y^2\rangle&=
\frac{ v_0^2\tau\gamma}{ k_y } \frac{ (1+\frac{\tau}{\gamma}k_y) }{(1+\frac{\tau}{\gamma}k_y)^2 +\Omega^2\tau^2} \,.
\end{flalign}
Expression~\eqref{eq:pxy_elliptic} shows that the interplay between chirality and elliptic confinement induces a cross-correlation $\langle xy\rangle$. 
The shape deformation of the probability distribution observed numerically  in Fig.~\ref{fig:fig3} is
described analytically by the formula:
\begin{equation}
\label{eq:crosscorrelation}
\langle xy\rangle=\omega\tau \frac{v_0^2\tau\gamma}{k_x+k_y}\Bigl( \frac{ 1 }{(1+\frac{\tau}{\gamma}k_y)^2+\omega^2\tau^2}
- \frac{ 1 }{(1+\frac{\tau}{\gamma}k_x)^2+\omega^2\tau^2} \Bigr) \,.
\end{equation}
The cross-correlation vanishes for $\omega \to 0$ and displays a non-monotonic behavior as a function of the reduced chirality: it is positive or negative depending on the sign of $\omega$ and on the ratio $k_y/k_x$,
and vanishes when the radial symmetry is restored ($k_x=k_y$).

As in the case of radial potential, the ABP dynamics displays a richer scenario (Fig.~\ref{fig:fig3}~(a)-(d)).
For small reduced chirality $\omega \tau \ll 1$ (Fig.~\ref{fig:fig3}~(a)), particles accumulate away from the potential minimum along the ellipsoid determined by the potential. In particular, particles accumulate more along the $x$ direction where the system is more confined, with respect to the $y$ direction.
In this regime, the increase of the chirality is able to change the orientation of the accumulation area introducing an evident asymmetry in the shape of $p(x,y)$ (Fig.~\ref{fig:fig3}~(b)).
This effect is enhanced when the reduced chirality is increased, until the regime $\omega \tau \sim 1$.
Correspondingly, the tendency of particles to climb on the potential is reduced and we can observe larger spatial fluctuations (Fig.~\ref{fig:fig3}~(c)).
The mechanism that leads to the latter effect is equal to that described in Sec.~\ref{sec:radial}.
Finally, spatial fluctuations are consistent (Fig.~\ref{fig:fig3}~(d)) as if the system was governed by a smaller effective temperature until the accumulation far from the potential minimum is completely suppressed.
Again, this is consistent with the results described for a chiral particle in a radial potential.

Both AOUP and ABP dynamics are characterized by a non-Maxwell-Boltzmann distribution with a breaking of the parity symmetry with respect to the $x$ (or $y$) axis that characterizes the elliptic potential. 
In other words, even if $U(-x, y)=U(x,)$, we have $p(-x,y)\neq p(x,y)$ (or equivalently $p(x,-y)\neq p(x,y)$).
This effect emerges in the occurrence of spatial correlations between the Cartesian components of the positions and is purely due to the interplay between chirality and asymmetry of the potential.

\begin{figure*}[t!]
\includegraphics[width=0.95\textwidth]{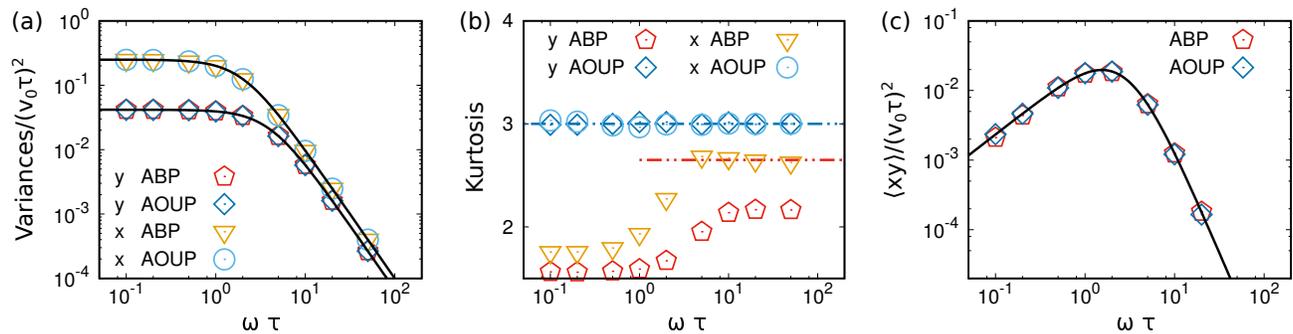}
\caption{
{\textbf{Moments and cumulants for the distribution of chiral active particles in a harmonic elliptic potential.}}
Panel (a): Variances of the distribution for $x$ and $y$ components of the distribution $p(x,y)$, i.e. $\langle x^2\rangle$ and $\langle y^2\rangle$, as a function of the reduced chirality $\omega \tau$.
Panel (b): Kurtosis of the distribution for $x$ and $y$ components, i.e. $\langle x^4\rangle/\langle x^2\rangle^2$ and $\langle y^4\rangle/\langle y^2\rangle^2$, as a function of $\omega\tau$. 
Panel (c): Cross-correlation $\langle x y \rangle$ as a function of $\omega \tau$.
In all the panels, results are presented both for ABP and AOUP dynamics as indicated in the legend.
The remaining parameters of the simulations are: $k_y/k_x=3$, $k \tau/\gamma=1$, and $D_t/(v_0^2 \tau)=0$.
}
\label{fig:fig4}
\end{figure*}

\subsection{Moments of the distribution}
To quantify this effect we consider the moments of the distribution for $x$ and $y$ coordinates (Fig.~\ref{fig:fig4}).
Specifically, Fig.~\ref{fig:fig4}~(a) displays the variances $\langle x^2 \rangle$ and $\langle y^2 \rangle$ as a function of the reduced chirality $\omega \tau$.
The results are similar for both ABP and AOUP and agree with the theoretical prediction Eq.~\eqref{eq:elliptic_theory_varx} and Eq.~\eqref{eq:elliptic_theory_vary}.
The variances of the distribution that can be interpreted as the effective temperature of the system decrease for both $x$ and $y$ components approximatively when $\omega\tau \approx 1$.
However, the effect of chirality manifests itself for smaller values of $\omega \tau$ when the system is less confined, i.e.\ along the $y$ component.
For $\omega \tau \gg 1$, the chirality decreases the effective temperature of the system as $\sim \omega^{-2}$.

Similarly to Fig.~\ref{fig:fig3}, to quantify the non-Gaussian nature of the system we study the kurtosis along $x$ and $y$ components, defined as $\langle x^4\rangle/\langle x^2 \rangle^2$ and $\langle y^4\rangle/\langle y^2 \rangle^2$.
In agreement with our intuition, the kurtosis of the AOUP model 
for every value of $\omega \tau$, is equal to $3$.
In the ABP case, the two kurtosis display the same qualitative behavior observed in the case of the radial potential in Sec.~\ref{sec:radial}. They start from values close to $2$, when the system displays accumulation far from the potential minimum, and then increase with $\omega \tau$, until reach an asymptotic value slightly smaller than $3$.
Here, the non-Gaussian nature of the chiral ABP is more evident along the $x$ axis when the system is more confined.

Finally, we plot the cross-correlation $\langle x y \rangle$, as a function of $\omega \tau$, where again, the ABP and AOUP display similar results.
The cross-correlation of both models is reproduced by the theoretical prediction~\eqref{eq:crosscorrelation} that shows a non-monotonic behavior.
In the regime of small reduced chirality, $\omega \tau \ll 1$, the cross-correlation starts from zero and then grows almost linearly until reaches a maximum around $\omega \tau \approx 1$.
From here, further increase of $\omega \tau$ reduces the value of $\langle x y \rangle$ with a scaling $\sim \omega^{-2}$ until vanishes.

\subsection{Conditional moments of the distribution}

\begin{figure*}[t!]
\includegraphics[width=1\textwidth]{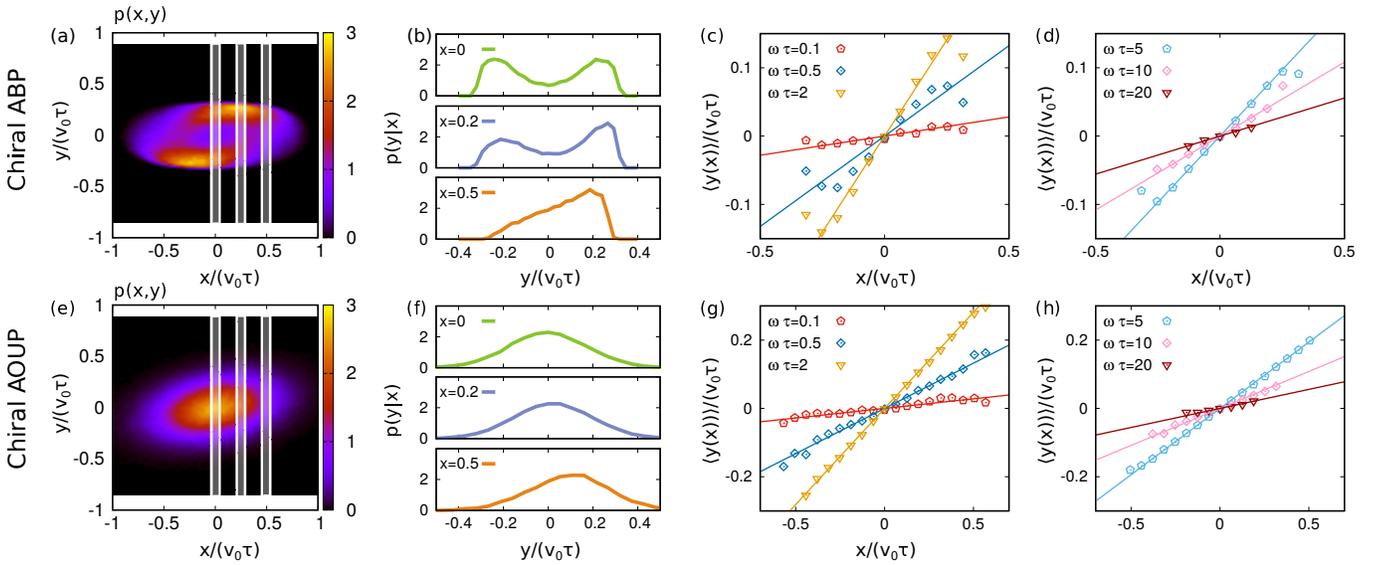}
\caption{
{\textbf{Conditional distribution and moments for chiral active particles in a harmonic elliptic potential.}}
Panel (a) and (e): distribution $p(x,y)$, as a function of the reduced positions $x/(v_0\tau)$ and $y/(v_0\tau)$ for ABP (a) and AOUP (e).
Panel (b) and (f): conditional probability distribution $p(y|x)$ (at fixed $x$) for three positions $x/(v_0\tau)=0, 0.2, 0.5$ for ABP (b) and AOUP (f). these positions are also marked in panels (a) and (b).
Panel (c), (d), (g), (h): first conditional moment, $\langle y(x)\rangle$, calculated at fixed $x/(v_0\tau)$, as a function of $x/(v_0\tau)$ for several values of the reduced chirality $\omega\tau$.
In these panels, colored points are obtained by simulations, while correspondingly colored solid lines are the theoretical predictions obtained by Eq.~\eqref{eq:cond_first_mom}.
The remaining parameters of the simulations are: $k_y/k_x=3$, $k \tau/\gamma=1$, and $D_t/(v_0^2 \tau)=0$.
}
\label{fig:fig5}
\end{figure*}

To underpin the breaking of the parity symmetry of the distribution induced by the interplay between chirality and potential asymmetry, we study the conditional distribution of the system, $p(y|x)$, i.e. the distribution calculated at fixed $x$, defined as $p(y|x)=p(x,y)/p_1(x)$ (Fig.~\ref{fig:fig5}) and the corresponding first conditional moment.
Fig.~\ref{fig:fig5}~(b) and (f) show $p(y|x)$ for $\omega\tau=2$ for three positions $x/(v_0\tau)=0, 0.2, 0.5$ considered as examples. Panel (b) refers to the ABP dynamics (whose joint distribution, $p(x,y)$, is reported in Fig.~\ref{fig:fig5}~(a)) while panel (c) refers to the AOUP dynamics (whose $p(x,y)$ is reported in Fig.~\ref{fig:fig5}~(e)).

In the AOUP case, the distribution has a Gaussian shape in all the cases.
However, for $x/(v_0\tau)=0$, the Gaussian is centered in the origin while by increasing $x/(v_0\tau)$, the center of the Gaussian shifts to values larger than zero.
In other words, the parity symmetry (characterizing the elliptic potential) is broken at fixed $x/(v_0\tau)$, i.e. $p(y|x)\neq p(-y|x)$.
This is consistent with our analytical prediction
\begin{equation}
p(y|x)= C'\, \exp\Bigl(-\frac{1}{2}\frac{\langle xy \rangle^2  x^2+\langle x^2 \rangle^2 y^2- 2\langle xy\rangle \langle x^2 \rangle  xy}{(\langle x^2 \rangle\langle y^2 \rangle-\langle xy \rangle^2) \langle x^2 \rangle}\Bigr)
\end{equation}
and
\begin{equation}
\label{eq:cond_first_mom}
\langle y(x) \rangle = \frac{\langle xy\rangle}{\langle x^2\rangle}x
\end{equation}
is the first conditional moment of the distribution, i.e. the average $y$ at fixed $x$, as a function of $x$.

As clear from the shape of $p(x,y)$ and known results in the absence of chirality, the ABP has a non-Gaussian distribution.
The conditional distribution of both models shows a similar degree of asymmetry and, in particular, the breaking of the parity symmetry in the distribution $p(y|x) \neq p(-y|x)$.
Indeed, at $x/(v_0\tau)=0$, the $p(y|x)$ displays a fully symmetric bimodal profile.
For larger values of $x/(v_0\tau)$, the spatial shape of $p(y|x)$ displays intrinsic asymmetry: the right peak of the distribution becomes larger than the left until the left peak is completely suppressed.

To characterize this asymmetry, we study the first conditional moment of the distribution $\langle y(x)\rangle$.
This analysis is reported in Fig.~\ref{fig:fig5}~(g) and (h) for the AOUP case and in Fig.~\ref{fig:fig5}~(c) and~(d) for the ABP dynamics for several values of the reduced chirality $\omega\tau$.
In both cases, $\langle y(x)\rangle$ is described by a linear profile with the same slope, in agreement with our theoretical prediction Eq.\eqref{eq:cond_first_mom}.
$\langle y(x)\rangle$ shows an almost flat profile for $\omega \tau \ll 1$, as expected from the non-chiral case.
The slope is an increasing function of the chirality until reaches a maximum for $\omega\tau=2$.
For larger values of $\omega\tau$, the slope decreases again until becomes almost flat.
This non-monotonicity explains the one observed in the behavior of the cross-correlation $\langle xy\rangle$ (Fig.~\ref{fig:fig4}~(c)). Indeed, the non-zero conditional moment $\langle y(x)\rangle$ induces global cross-correlations in the full distribution and thus, the larger $\langle y(x)\rangle$, the larger $\langle xy\rangle$.

\section{Conclusions}\label{sec:conclusions}

In summary, we have studied a chiral active particle confined in an external potential, with and without radial symmetry.
For radial potentials, the chirality affects the effective temperature of the system both for ABP and AOUP dynamics.
Specifically, in the AOUP case, the dynamics displays Gaussian properties due to the linearity of the system with an effective variance that decreases with the chirality.
In the ABP case, the chirality reduces the non-Gaussianity of the system, by suppressing the accumulation far from the minimum of the potential typical of  the non chiral confined ABP. In other words, the chirality induces a transition from a bimodal to a unimodal density.

For non-radial potentials, the scenario is richer due to the interplay between chirality and asymmetry of the potential which is able to break the parity symmetry in the probability distribution of the system. 
As a consequence, a non-Maxwell-Boltzmann distribution is found both for chiral ABP and chiral AOUP dynamics.
This effect emerges in cross-correlations between the Cartesian components of the position that are present both for chiral ABP and chiral AOUP.
The linearity of the AOUP makes possible analytical calculations that allow us to analytically predict the first two moments of the chiral ABP in a harmonic potential.


\acknowledgments
LC acknowledges support from the Alexander Von Humboldt foundation.
HL acknowledges support by the Deutsche Forschungsgemeinschaft  (DFG)  through  the  SPP  2265,  under  grant  number LO 418/25-1.

\appendix
\section{Derivation effective equation for the probability distribution function}
\label{sec:appendixA}

Although the linear models can be solved by considering the Langevin equation for the coordinates and then deriving the distribution function from the first non vanishing cumulants, an equivalent description is possible in terms of an effective
Fokker-Planck equation (FPE) for the distribution function. At the linear level, the two methods yield equal results and the choice between them is a matter of taste, but when the potential is non quadratic the FPE method is simpler to implement. 

Here, we develop the second method in the case of chiral active particles. 
For the sake of completeness, we briefly illustrate the basic assumptions leading to a closed  equation for the probability density distribution~\cite{fox1986functional,hasegawa2007moment,rein2016applicability}. 
The equation of motion~\eqref{eq_t1} for $D_t=0$ can be written for each component as
\begin{eqnarray}
\gamma \frac{d   \bar x_m(t)}{dt}\!\!&=&\!\! F_m +  \eta_m(t) 
\label{eq:overdampedmotion}
\end{eqnarray}
where the index $m$ marks denotes different Cartesian components (for instance, $m=x,y$ in two dimensions) and $\eta_m$ is a component of the active force $\gamma v_0 \mathbf{n}$.
By standard manipulations, we derive the equation for the associated probability distribution function
\begin{eqnarray}\nonumber\\&&
\frac{\partial}{\partial t} p(\{x\},t) =
- \frac{1}{\gamma}\sum_m\frac{\partial}{\partial x_m} F_m(\{x\}) p(\{x\},t)
\nonumber\\&&
- \sum_{m} \frac{\partial}{\partial x_m} 
\langle \eta_m(t) \hat \rho(\{x\},t)\rangle \,.
\label{eq:masterequationb}
\end{eqnarray}
where $\hat \rho(\{x\},t)= \Pi_m\delta(\bar x_m(t)-x_m)$, with $x_m$ the local value assumed by $\bar{x}_m$, and $p(\{x\},t)= \langle \hat \rho(\{x\},t)\rangle$. 
The average $\langle \cdot\rangle$ is performed over the  realizations of the stochastic process $\eta_m$ and the curly brackets are used to denote a dependence over all the components of a vector.

Since Eq.~\eqref{eq:masterequationb} is not a closed equation for the probability distribution function,
we employ the Novikov formula~\cite{novikov1965functionals} to evaluate the average appearing in the last term. 
This formula is valid for arbitrary Gaussian random functions (Note that the ABP is not described by a Gaussian noise):
\begin{equation}
\langle \eta_m(t) R[\{\eta\}] \rangle
=\int_0^t \:dt' \:\sum_n C_{mn}(t,t')  
\left< \frac{\delta R[\{\eta\}] }{\delta \eta_n} \right>	
\label{has5}
\end{equation}
where $R[\{\eta\}] $ denotes a functional of $\{\eta\}$ and on the right hand side is the variational derivative of this functional. 
The term 
\begin{eqnarray}
C_{mn}(t,t')=
\langle\eta_m(t) \eta_n(t') \rangle
\label{has6}
\end{eqnarray}
 is the active force correlation function.
Employing Eq.~\eqref{has5}  and the definition of $\hat \rho(\{x\},t)$, we get
\begin{eqnarray} \nonumber\\&&
 \langle \eta_m(t) \hat \rho(\{x\},t)\rangle 
\\&&=  \int_0^t \:dt' \:\sum_n C_{mn}(t,t')
\sum_k \left< \frac{\delta( \hat \rho(\{x\},t)) }{\delta \bar x_k} 
\frac{\delta \bar x_k(t)}{\delta \eta_n(t')} \right> \nonumber
\\&&
= - \sum_k \sum_n \frac{\partial }{\partial x_k} \int_0^t \;dt' \:C_{mn}(t,t')
\left< \hat \rho(\{x\},t)
\frac{\delta \bar x_k(t)}{\delta \eta_n(t')} \right>\nonumber\,.
\label{has7}
\end{eqnarray}
The functional derivative of $\bar x_k(t)$ with respect to $\eta_n(t')$ is given by the following expression valid for for $t>t'$
\begin{eqnarray}
\frac{\delta \bar x_k(t)}{\delta \eta_n(t')}&=&  \theta(t-t')
\left[\exp \int_{t'}^{t}\:ds \:  {\bf J}(s)\right]_{kn} 
\label{has13}
\end{eqnarray}
where the matrix ${\bf J}(s)$ has elements $ J_{kl}(t) =\frac{1}{\gamma} \frac{\partial F_k(\{x(t)\})}{\partial x_l(t)}$.
Combining Eq.~\eqref{has7} with Eq.~\eqref{has13}, we find
\begin{eqnarray}
\label{eqappendix:average}
&&\langle \eta_m(t) \hat \rho(\{x\},t) \rangle
= 
- \sum_k \sum_n \frac{\partial}{\partial x_k} \\
&&
\int_0^t dt'  \left[C_{mn}(t,t') \left< \hat \rho(\{x\},t)
\left (\exp \int_{t'}^{t} \:ds  {\bf J}(s) \right)_{kn} \right> \right] \,.  
\label{has14}
\nonumber 
\end{eqnarray}
The expressions obtained up to here are exact but not close.
Therefore, we employ a closure scheme to obtain a theoretical prediction for the probability distribution.
To achieve this goal, we estimate the Eq.~\eqref{eqappendix:average} as follows:
\begin{eqnarray}&&
 \left< \hat \rho(\{x\},t)
\exp\left(\int_{t'}^{t} \: ds  {\bf J}(s) \right)_{kn} \right> \nonumber\\&&
\simeq 
 \Bigl<  \hat \rho(\{x\},t) \Bigr>
 \left( \exp \Bigl<     {\bf J}(t)    \Bigr> (t-t') \right)_{kn}  \,.
\label{has15}
\end{eqnarray}
Here, we have performed three approximations: 1) the factorization of the averages; 2) the replacement of the average of the exponential with the exponential of the average. 3) we have treated ${\bf J}(s)$ as a constant in the time integral in the exponent.
Let us remark that the above approximations are exact in the case of quadratic potentials because $\mathbf{J}(t)=\text{const}$ and not an approximation as in the general case.
Going back to Eq.~\eqref{has7}, we find
\begin{eqnarray}
&&\langle \eta_m(t) \hat \rho(\{x\},t) \rangle
= - \sum_k  \frac{\partial}{\partial x_k}     p(\{x\},t)  D_{mk}(t) 
\label{has17}
\end{eqnarray}
where we have defined the following  matrix elements:
\begin{eqnarray}
&&
D_{mk}(t)= \sum_n \Bigl[\int_0^t d\tilde t  C_{mn}(\tilde t)
 \left(  \exp\Bigl<     {\bf J}(t)     \Bigr>  \tilde t \right)_{nk} \Bigr] 
\label{has18c}
\end{eqnarray}
Finally, we obtain a closed equation for the probability distribution
\begin{eqnarray}&&
\frac{\partial}{\partial t} p(\{x\},t) = \\&&
-  \sum_m\frac{\partial}{\partial x_m} \frac{F_m(\{x\})}{\gamma} p
 + \sum_{m k} \frac{\partial}{\partial x_m} \Bigl[
  \frac{\partial}{\partial x_k}  D_{mk}   p
\Bigr] \nonumber\,.
\label{has4b}
\end{eqnarray}
%
The method developed here (and in particular the approximations 1), 2) and 3) in Eq.~\eqref{has15}) are exact in the case of a chiral AOUP particle confined in a harmonic potential with radial or non-radial (elliptic) shape.
In constrast, for non-linear forces, 1), 2) and 3) are approximations whose accuracy depends on the potential considered.
Finally, the method represents only an approximation for the ABP because the Novikov formula, Eq.~\eqref{has5}, does not hold. Indeed, the ABP is governed by a non-Gaussian noise because $\mathbf{n}$ is an orientation with a non-fluctuating unit modulus. 

\section{Application to simple cases.}
\label{sec:appendixB}

The general method presented in the previous appendix is applied to a confining potential (with radial and non-radial symmetry) studied in Sec.~\ref{sec:radial} and Sec.~\ref{sec:nonradial}.
First, we estimate the components of the time-autorocorrelation of the active force $C_{mn}(t-t')$:
\begin{eqnarray}&&
C_{mn}(t-t') = \nonumber\\&&v_0^2  \, e^{-|t-t'|/\tau}
\left(\begin{array}{ccccccc}
\cos(\omega (t-t')) &    -  \sin(\omega |t-t'|)\\   
 \sin(\omega |t-t'|) &  \cos(\omega (t-t'))
\end{array}\right)  \,.
\label{eq:activeforcecorrelation}
\end{eqnarray}
Then, we estimate $D_{mk}(t)$ for a rather general form of central potential, $U(r)$, applying the definition~\eqref{has18c}
and taking the limit $t\to \infty$. 
We obtain the following matrix elements
\begin{eqnarray}&&
D_{xx}=
 \frac{v_0^2 \tau}{r^2}\Bigl[y^2 u_{I}-xy w_{I} +
 x^2 u_{II}+xy w_{II} \Bigr] \\&&
 D_{yy}= \frac{v_0^2 \tau }{r^2}\Bigl[x^2 u_{I}+xy w_{I} +
 y^2 u_{II}-xy w_{II} \Bigr]\\&&
 D_{xy}= \frac{v_0^2 \tau }{r^2} 
   \Bigl[ w_{I} x^2-xy  u_{I}    +
 w_{II} y^2 + xy u_{II} \Bigr]\\&&
D_{yx}= -\frac{v_0^2 \tau}{r^2} 
   \Bigl[ w_{I} y^2+xy   u_{I}      +
 w_{II}  x^2 -xy u_{II} \Bigr]
\end{eqnarray}
where  we used the abbreviations:
\begin{eqnarray}&&
u_{II}= \frac{  (1+\tau \frac{U''}{\gamma})       }{(1+\tau \frac{U''}{\gamma})^2+\omega^2\tau^2}\\&&
u_{I}= \frac  {(1+\tau \frac{U'/r}{\gamma})       }{(1+\tau\frac{U'/r}{\gamma})^2+\omega^2\tau^2}\\&&
w_{II}= \frac{  \omega  \tau    }{(1+\tau \frac{U''}{\gamma})^2+\omega^2\tau^2}\\&&
w_{I}= \frac  {\omega   \tau   }{(1 +\tau \frac{U'/r}{\gamma})^2+\omega^2\tau^2}\,.
\end{eqnarray}
and the primed symbols stand for the first and second derivatives of $U(r)$.
After eliminating $x$ and $y$ in favor of the radial coordinate $r=\sqrt{x^2+y^2}$,
the resulting effective Fokker-Planck equation is conveniently written as:
\begin{eqnarray}&&
\frac{\partial}{\partial t} p=  
 \frac{1}{r}\frac{\partial}{\partial r} \Bigl[ r\frac{ U'(r)}{\gamma} p 
 + v_0^2\tau  \Bigl(    (u_{II} -      u_{I} )p+
    r \frac{\partial}{\partial r}  ( u_{II} p)
 \Bigr)\Bigr] \,.\nonumber\\&&
\label{has4bcd4}
\end{eqnarray}
The time independent solution of Eq.~\eqref{has4bcd4} is obtained by imposing the vanishing of the radial
component, ${\cal J}_{rad}$, of the probability current (i.e. minus the expression contained in the square parenthesis in the r.h.s. of Eq.~\eqref{has4bcd4}). 
For the particular case where  is harmonic ($U(r)=kr^2/2$), expression~\eqref{has4bcd4}, the difference 
$(u_{II} -      u_{I} )$ vanishes and the explicit solution is:
\begin{equation}
p(r)=\rho_0 \exp\Bigl( - \frac{(1+\frac{\tau}{\gamma} k)^2+\omega^2\tau^2}{ (1+\frac{\tau}{\gamma}k)    } \,\frac{1}{\tau \gamma v_0^2} \frac{k r^2}{2}
 \Bigr) \, ,
 \end{equation}
 while for arbitrary central potentials the problem can always be reduced to a simple quadrature.
 Interestingly, it is easy to verify
that due to the handedness of the system the tangential component  of the probability current does not vanish whenever $\omega\tau\neq 0$. In other words, the presence of a radial gradient
in the probability density induces a circulation of the particles in the direction orthogonal to it, but such a current does not affect the probability distribution itself. The tangential current reads:
 \begin{eqnarray}&&
 {\cal J}_{tan}=  v_0^2\tau  \frac{ \omega\tau   }{(1+\frac{\tau}{\gamma}k)^2+\omega^2\tau^2}  \frac{\partial}{\partial r}  \, p(r)
 \end{eqnarray}
By expressing $p(r)$ as a function of the Cartesian components we obtain Eq.~\eqref{eq:p(x,y)theory}.

By contrast , in the case of the elliptic quadratic confining potential, $U(r)=(k_x x^2+k_y y^2)/2$, 
one cannot exploit the radial symmetry of the problem and 
the equation for the probability density reads:
 \begin{eqnarray}
&&
  \frac{\partial}{\partial t}   p(x,y,t) = \frac{\partial}{\partial x} \frac{k_x x}{\gamma} p  +\frac{\partial}{\partial y}\frac{k_y y}{\gamma}  p \\
&&
+  v_0^2\tau \Bigl[ \frac{ (1+\frac{\tau}{\gamma}k_x)    }{(1+\frac{\tau}{\gamma}k_x)^2+\omega^2\tau^2} \frac{\partial^2}{\partial x^2}  
  +  \frac{ (1+\frac{\tau}{\gamma}k_y)    }{(1+\frac{\tau}{\gamma}k_y)^2+\omega^2\tau^2} \frac{\partial^2}{\partial y^2}  
  \nonumber
\\
&&    + (  \frac{ \omega\tau    }{(1+\frac{\tau}{\gamma}k_y)^2+\omega^2\tau^2}      -  \frac{ \omega\tau    }{(1+\frac{\tau}{\gamma}k_x)^2+\omega^2\tau^2}) \frac{\partial^2}{\partial x\partial y} \Bigr]  p \, .\nonumber
\label{ellipticprob}
 \end{eqnarray}
 The steady probability $p(x,y)$ can be obtained by first determining its cumulants (Eqs.~\eqref{eq:elliptic_theory_varx},~\eqref{eq:elliptic_theory_vary},~\eqref{eq:crosscorrelation})
 from Eq.~\eqref{ellipticprob} and 
 using this information to express the pdf as in Eq.~\eqref{eq:p(x,y)theory}.


\bibliographystyle{rsc} 

\bibliography{SD25maggio.bib}


\end{document}